\documentclass[12pt]{iopart}
\usepackage{color}
\usepackage{graphicx}
\usepackage{subfigure}
\usepackage{lineno}

\usepackage{latexsym,amsmath,verbatim}
\usepackage{iopams}

\begin{document}

\title{Rounding of abrupt phase transitions in brain networks}

 \author{Paula Villa Mart{\'\i}n, Paolo Moretti, and Miguel A. Mu\~noz}
\address{Departamento de Electromagnetismo y F\'\i sica de la
  Materia and Instituto Carlos I de F\'\i sica Te\'orica y
  Computacional, Facultad de Ciencias, Universidad de Granada, 18071
  Granada, Spain }

\begin{abstract} 
  The observation of critical-like behavior in cortical networks
  represents a major step forward in elucidating how the brain manages
  information.  Understanding the origin and functionality
  of critical-like dynamics, as well as their robustness, is a major
  challenge in contemporary neuroscience.  Here, we present an
  extensive numerical study of a family of simple dynamic models,
  which describe activity propagation in brain networks through the
  integration of different neighboring spiking potentials, mimicking
  basic neural interactions. The requirement of signal integration may
  lead to discontinuous phase transitions in networks that are well
  described by the mean-field approximation, thus preventing the
  emergence of critical points in such systems. Here we show that
  criticality in the brain is instead robust, as a consequence of the
  hierarchical organization of the higher layers of cortical networks,
  which signals a departure from the mean-field paradigm. We show
  that, in finite-dimensional hierarchical networks, discontinuous
  phase transitions exhibit a rounding phenomenon and turn continuous
  for values of the topological dimension $D\le 2$, due to the
  presence of structural or topological disorder.  Our results may
  prove significant in explaining the observation of traits of
  critical behavior in large-scale measurements of brain activity.
\end{abstract}



\maketitle


\section{Introduction}
Experimental evidence of critical or quasi-critical behavior in brain
networks was gathered over the past decade
\cite{BP03,BP04,Plenz07,Beggs08,Peterman09,Haimovici,Chialvo10}.  The
discovery of scale-invariant avalanches of neural activity led to the
conjecture that the brain might operate close to a critical point
\cite{BP03,BP04}.  It was argued that critical behavior might bear
functional advantages; for instance, the divergence at criticality of
quantities such as susceptibilities and correlation lengths could
entail the ability of brain networks to coordinate system-wide
activities and efficiently respond to a broad range of stimuli.  A
vast number of studies have since flowered, focusing on the numerical
simulation of simple dynamical models that could recover
phenomenologically the hallmarks of criticality observed in
experiments \cite{Kaiser07,Kaiser10,Rubinov,Zhou11,Zhou12}.  In
particular, it was noted that effective highly-simplified models of
activity propagation --such as the contact process and the
quiescent-excited-refractory-quiescent model-- could provide valuable
information on large-scale brain properties
\cite{Grinstein-Linsker}. In these ideal models, an active ``unit'' or
node --be it a neuron at a microscopic scale or a coarse-grained
active region at a larger mesoscopic scale-- can propagate its
activity to neighboring units and/or become deactivated.  Such simple
dynamics --where activity propagation involves a single active node--
lead generically to continuous phase transitions, with a critical
point separating an active from a quiescent phase
 \cite{Marro-Dickman,Liggett,Odor-book,Hinrichsen-book}. Moreover,
relatively simple modifications of these models implementing standard
mechanisms of self-organized criticality lead to robust critical or
quasi-critical behavior without the need of parameter fine tuning
\cite{Levina07,Mill10,JaboJSTAT}.

However, a closer look at real neural dynamics suggests that neural
activity propagation may follow more complicated rules. In particular,
individual neurons usually require to integrate up to hundreds of
post-synaptic potentials before spiking themselves, as typically
captured by integrate-and-fire models
\cite{Abbott,Burkitt-I,Burkitt-II}. At mesoscopic scales, such
requirement may be less stringent; however it is reasonable to
consider that a few neighboring active units might be required to
generate further activity: i.e. the dynamics follow a schematic rule of 
the type: $n$A$\to (n+m)$A, with $n>1$ and $m$ of the order of a few
units, where each A stands for an active location or site
\cite{Marro-Dickman,Liggett,Odor-book,Hinrichsen-book}.

Such types of $(n,m)$-processes are well known in reaction-diffusion
systems \cite{Marro-Dickman,Liggett,Odor-book,Hinrichsen-book}, and
they are often used in the modeling of neural dynamics. In particular,
it is known that they lead to broad phases of sustained activity
\cite{Kaiser07} and enhanced dynamic ranges
\cite{Kinouchi-Copelli,Gollo12}.  However, these processes are also
well-known to lead to pattern formation (Turing patterns) and to
discontinuous phase transitions between active and quiescent
phases, with associated phase coexistence \cite{Windus-Jensen07} and
lacking critical points, in seeming contradiction with the observation
of scale-invariant behavior in brain dynamics. In other words, the
requirement of more than one source of activity to generate further
activity leads to discontinuous phase transitions, separating two
highly different active and quiescent phases, with no sign of
criticality nor scale-invariance in between
\cite{Marro-Dickman,Liggett,Odor-book,Hinrichsen-book}.

Our goal here is to reconcile the need for signal integration at the
neuron scale, supposedly leading to discontinuous transitions, with
the empirical observation of critical-like features, characteristic of
continuous phase transitions. As we hope to convincingly argue, the
key to this puzzling ambiguity lies in the topology of the underlying
network of neural connections, which we will prove responsible for the
generic rounding of discontinuous transitions. In a nutshell, and in
analogy with what happens in problems of thermodynamic equilibrium at
low dimensions, the existence of some form of structural disorder
implies that potentially discontinuous transitions are rounded-off, thus making
the system critical.

\section{The underlying network of neural connections}

In spite of the huge complexity that would be required to represent
the detailed structure of the brain down to the single-neuron level,
an effective coarse-grained description of neural contact patterns can
be provided by a network --the connectome-- whose nodes represent
groups of neurons, such as cortical columns, and whose links represent
the groups of fibers connecting them \cite{Kaiser10}.

Studies employing different neuroimaging techniques have revealed that
the Human Connectome, the current mapping of human brain connections,
is organized in a hierarchical and modular fashion, in which local
regions are clustered into large-scale moduli, which in turn form
higher level structures and so on
\cite{Sporns05,Meunier,Kaiser07b,Sporns07,Hagmann,Zamora-Lopez}.  The
resulting hierarchical modular network (HMN) can be visualized as
built-up from moduli of large internal neural connectivity, enclosed
into higher-level sparser moduli, in a nested hierarchical fashion.

HMNs have been recently found to play a crucial role in neural
dynamics. In particular, simple models of activity propagation were
recently found to display Griffiths phases when running on top of HMNs
\cite{NatCommun}, corroborating the experimental observation of
extended critical regions in the human at its resting state
\cite{Taglia}. Similarly, they were argued to extend the region of
apparent criticality in self-organized models of neural activity and
they were used to explain the ability of the brain to sustain activity
over extended time windows
\cite{Kaiser07,Kaiser10,Rubinov,Zhou11,Zhou12,NatCommun}.

\begin{figure}
\begin{center}
\vspace{-0cm}
\includegraphics[width=12cm]{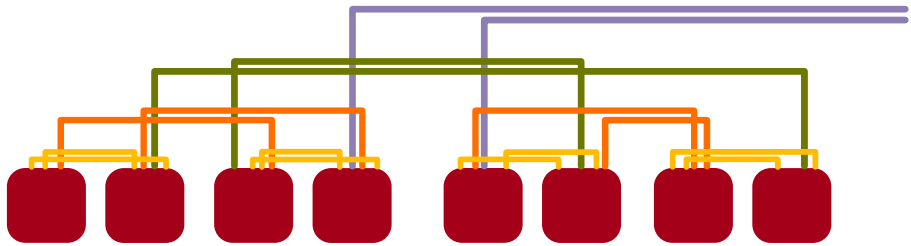}
\caption{Sketch of the HMN construction method. Given a positive
  integer $s$, consider $2^s$ basal fully connected moduli of size
  $M$. At the lowest hierarchical level, moduli are linked pairwise
  into super-moduli by establishing a fixed number $\alpha$ of random
  unweighted and undirected links between the elements of each modulus
  ($\alpha = 2$ in figure). Newly formed blocks are then iteratively
  linked pairwise with the same $\alpha$ for a total of $s$
  iterations, until the network becomes connected. The resulting
  network has size $N=2^s M$.}
\label{fig:network}
\end{center}
\end{figure}

Here we shall use a simple structural model to build-up synthetic HMNs 
as follows: local densely connected moduli are used as building blocks; 
they are recursively grouped by establishing additional inter-moduli links in a 
level-dependent way, as exemplified in Figure \ref{fig:network}.
Further details of the construction methods can be found in Reference
\cite{NatCommun}. A crucial feature of HMNs is represented by their
finite topological dimension $D$. The topological dimension of a
network can be defined as follows: starting from a single node, the
number of neighbors $N_z$ reachable after $z$ steps is computed for
increasing $z$ until the entire network is covered \cite{MAM-GP}. The
network is finite dimensional with dimension $D$ if $\langle
N_z\rangle\sim z^D$, generalizing the familiar behavior of regular
lattices. The topological dimension of a HMN can be tuned easily, by
changing the average number $\alpha$ of links between pairs of modules
at each hierarchical level (see Fig. \ref{fig:network} and
\cite{NatCommun}).  Although brain moduli and columns may be densely
connected, at larger mesoscopic and macroscopic scales the
hierarchical contact patterns become very sparse. At such scales, the
effective network becomes finite dimensional \cite{Gallos,NatCommun}.

\section{Continuous versus discontinuous transitions in the presence of
disorder}
It was recently conjectured that dynamical models of activity
propagation characterized by discontinuous phase transitions at the
mean-field level exhibit a rounding phenomenon in finite dimensional
disordered systems, eventually leading to continuous phase transitions
at dimension $D \le 2$ \cite{Villa14}.  Such behavior has been
envisaged as the non-equilibrium analogue of the well-know Imry-Ma
criterion, which states that --in the presence of quenched disorder--
spontaneous symmetry breaking as well as first-order phase transitions
are prohibited in equilibrium systems at $D \le 2$
\cite{Imry-Ma,Aizenman,Berker}.  Analogously to quenched disorder in
lattices, structural disorder is integral to HMNs as defined above and
may thus be responsible for the rounding of discontinuous phase
transitions in such systems \cite{MAM-GP}. In this light, we
investigate how this form of topological disorder can potentially
alter the order of phase transitions exhibited by simple
$(n,m)$-models of neural activity, which would normally exhibit first-order 
phase transitions in networks well described by the mean-field
approximation. In analogy with the Imry-Ma criterion, {\it a priori},
this effect should be expected to occur in networks with topological
dimension $D$ less than $2$ \cite{Villa14}.

\section{Results}
In what follows, we provide extensive numerical tests of the above
conjecture, showing how the topological dimension of a disordered
network can tune the nature of the dynamical phase transition,
ultimately forcing $n$A$\to (n+m)$A dynamics to exhibit continuous
transitions for $D \le 2$. To this end, we consider a prototypical
model, in which we choose $n=2$ and $m=1$ (a $(2,1)$-process in our
notation), whose Monte Carlo implementation is as follows: each of the
$N$ nodes of the network is endowed with a binary state variable
$\sigma=0,1$, inactive or active, and $\rho(t)$ is the density of
active nodes at time $t$; i) at each time step an active node
$\mathrm{R}_1$ is selected and time is increased by $[N\rho(t)]^{-1}$;
ii) with {\it death} probability $p_{\mathrm{d}}$, $\mathrm{R}_1$ is
deactivated, while with complementary probability $1-p_{\mathrm{d}}$,
a neighboring node $\mathrm{R}_2$ is considered and one of the
following actions is taken; iiia) if $\mathrm{R}_2$ is inactive,
activity diffuses to $\mathrm{R}_2$, leaving $\mathrm{R}_1$; iiib) if
$\mathrm{R}_2$ is active, a new neighbor $\mathrm{R}_3$ of
$\mathrm{R}_1$ is considered and, if inactive, it is activated with
{\it birth} probability $p_{\mathrm{b}}$.  From a neurophysiological
perspective, $p_{\mathrm{d}}$ encodes the exhaustion mechanism that
accounts for spontaneous deactivation of neurons and neural regions,
which proves essential in maintaining sustained activity bounded
\cite{Kaiser07}. The integrated activation is tuned by
$p_{\mathrm{b}}$. A simple mean-field equation for this type of dynamics is
\begin{equation}
 \dot{\rho}(t) = 
 -p_{\mathrm{d}}\rho(t) 
 +(1-p_{\mathrm{d}})p_{\mathrm{b}} \, \rho^2(t) \, [1-\rho(t)]
\label{mf}
\end{equation}
which exhibits a discontinuous phase transition (fold bifurcation) at
$p_{\mathrm{d}} = p_{\mathrm{b}}/(4+p_{\mathrm{b}})$
\cite{Windus-Jensen07}.  Notice that different choices of $n,m>1$,
which might account for enhanced realism in the physiological
description of brain networks at the mesoscale, would not affect such
behavior. On the other hand, a more detailed theoretical description
of this system --taking explicitly into account the underlying network
topology-- would be provided by a quenched-mean-field approach
\cite{Castellano10,Mata13}.

\begin{figure}
\begin{center}
\vspace{-0cm}
   \includegraphics[width=12cm]{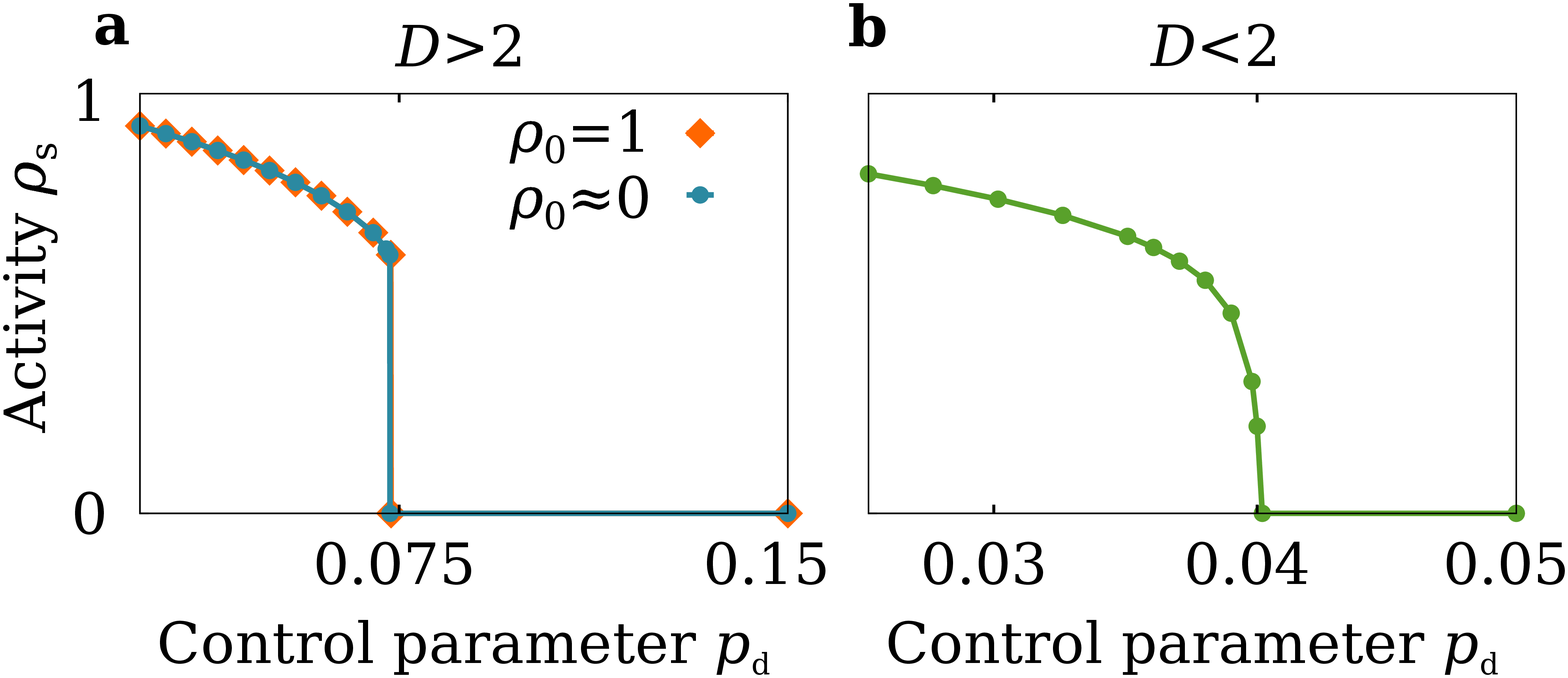}
   \caption{ (Color online) Phase diagrams for topological dimension
     $D$ above and below $D=2$ respectively. In high dimension, the
     phenomenology of a discontinuous phase transition is recovered,
     in agreement with the mean-field prediction for the
     $(2,1)$-process. Below $D=2$, the transition becomes continuous.
     A feeble appearance of hysteretic behavior is recorded in the
     $D>2$ case, where different colors correspond to different
     initial conditions and spinodal points marking the transition are
     located at $p_{\mathrm{d}_{\mathrm{thr}}}\approx 0.0732(1)$ and
     $p_{\mathrm{d}_{\mathrm{thr}}}\approx 0.0734(1)$ for $\rho_0=0$
     and $\rho_0=1$ respectively (not distinguishable in figure).
     Such dependence disappears for $D<2$, in accordance with the the
     hypothesis of a continuous phase transition, the critical point
     being located at $p_{\mathrm{d}_{\mathrm{thr}}}\approx 0.0402(1)$
     Simulations are run on HMNs of size $N=2^{17}=131072$,
     partitioned into $s=13$ hierarchical levels.}
   \label{fig:phasediagrams}
\end{center}
\end{figure}
As a substrate on which the above dynamics run, we considered
different HMNs, characterized by different topological dimensions $D$.
In the rest of the paper we will show results for HMN extracted from
two ensembles $\mathcal{N}_-$ and $\mathcal{N}_+$, each with a fixed
average dimension $D_{-,+}$ below and above the threshold value
$D=2$. In particular, we show results for $D_{-}\approx 1.6$ and
$D_{+}\approx 2.8$ (networks with such properties are obtained by
choosing $\alpha=1$ and $\alpha=4$ respectively, in the HMN
building-up process; see Figure \ref{fig:network}). 

We ran Monte Carlo simulations of the above $(2,1)$-process on such
networks.  Figure \ref{fig:phasediagrams} shows the steady state value
of the average activity density, as a function of the control
parameter $p_{\mathrm{d}}$, respectively below and above dimension
$D=2$, in spreading simulations starting both from localized active
seeds ($\rho_0\approx 0$) and from the homogeneously active state
($\rho_0=1$). While the discontinuity encountered above $D=2$ is in
agreement with the mean-field behavior for this type of dynamics,
below $D=2$ the dynamical phase transition is evidently continuous,
confirming the conjecture of a low-dimensional rounding.
\begin{figure}
\begin{center}
\vspace{-0cm}
   \includegraphics[width=12cm]{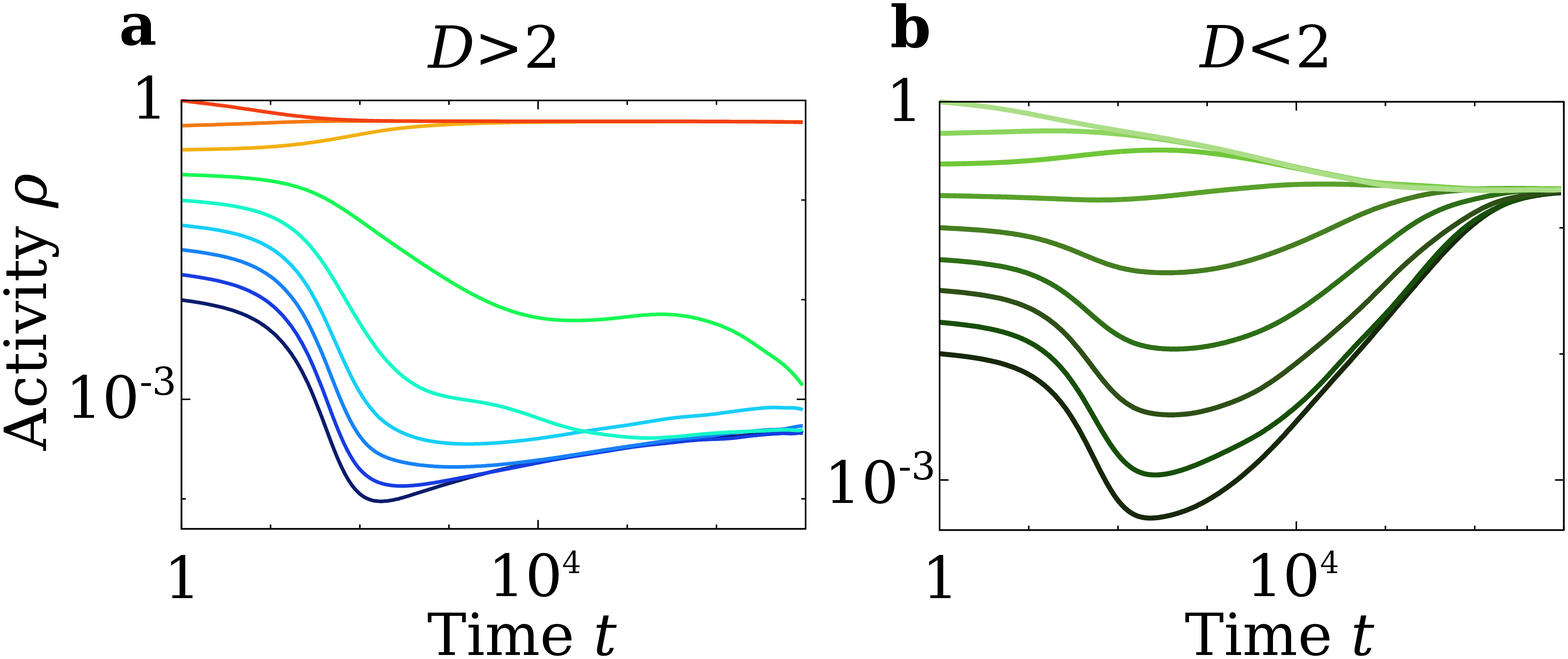}
   \caption{(Color online) Time evolution of the activity density
     $\rho$ for different initial values $\rho_0$ (different colors),
     for dimension above and below $D=2$ respectively.  In both cases, the
     control parameter $p_{\mathrm{d}}$ is chosen at the threshold value.  For
     high dimension, bistable behavior is recovered, as in standard
     first-order transitions, whereas no sign of bistability is
     encountered below $D=2$.  Notice however that both configuration
     converge very slowly to their expected behavior. In particular,
     in the $D>2$ (discontinuous) case, large enough initial
     conditions lead to very long transients, which could be
     misinterpreted as continuous behavior for short simulation
     times. Such traits of quasi-critical states become stronger as
     $D=2$ is approached from above, and corroborate the picture of a
     rounding phenomenon.}
    \label{fig:bistability}
\end{center}
\end{figure}

To provide further evidence of the radical difference in the
transition nature, Figure \ref{fig:bistability} shows the time
evolution of the average activity density $\rho$ upon changing the
initial activity $\rho_0$, for both cases in Figure
\ref{fig:phasediagrams}, each at the estimated threshold
$p_{\mathrm{d}}$. Above $D=2$ clear signs of bistability emerge,
signaling coexistence phenomena, which typically characterize
discontinuous phase transitions. Below $D=2$, however, the steady
state does not depend on the initial condition anymore, as expected
for a continuous transition, in which correlations become system-wide
and coexistence is prohibited. In order to gain a deeper understanding
of the rounding phenomenon, we can analyze the nature of the inactive
(absorbing) phase in both cases.
\begin{figure}
\begin{center}
\vspace{-0cm}
   \includegraphics[width=12cm]{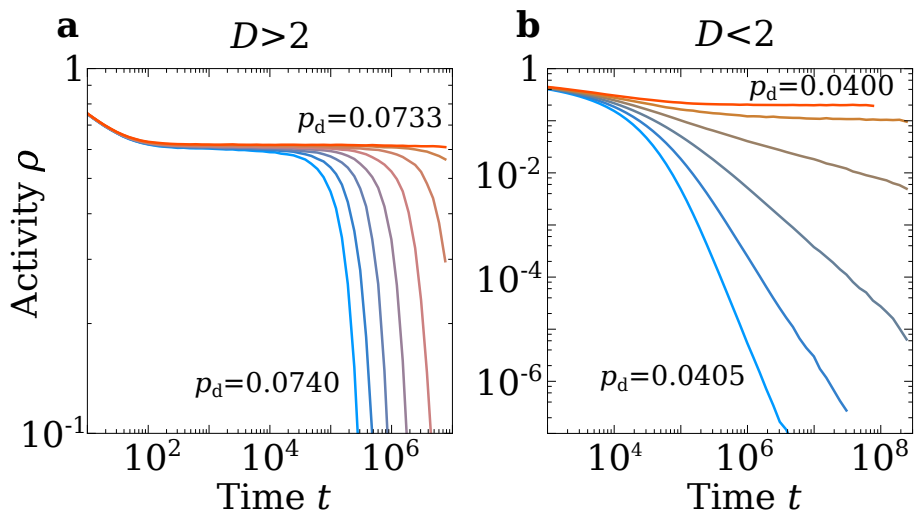}
   \caption{(Color online) Time evolution of the activity density
     $\rho$ for different values of the control parameter $p_{\mathrm{d}}$
     (different colors) below the dynamic threshold, for dimension above and
     below $D=2$ respectively.  For high dimension, the transition
     between the inactive and the active state is abrupt, with no
     signs of criticality. Below $D=2$, a Griffiths phase emerges
     for $p_{\mathrm{d}}>p_{\mathrm{d}_{\mathrm{thr}}}$, characterized by 
     generic
     power-law relaxation and critical-like behavior.}
   \label{fig:decay}
\end{center}
\end{figure}
To this end, let us consider simulations starting from a homogeneous
$\rho_0=1$ state. 
Time evolution of $\rho$ is shown in Figure \ref{fig:decay}, for
different values of $p_{\mathrm{d}}$ in the inactive phase.  
As usual, above $D=2$ results
for $p_{\mathrm{d}}$ point to an abrupt change in behavior at the dynamic
threshold $p_{\mathrm{d}_{\mathrm{thr}}}$, below which activity dies off
exponentially fast as soon as a large enough fluctuation breaks the 
coexistence of active and inactive islands. At
$p_{\mathrm{d}_{\mathrm{thr}}}$, such coexistence becomes stable in the
large-$N$ limit and the phenomenology of a discontinuous phase
transition is recovered. For dimensions below $D=2$, instead, the
system displays a Griffiths phase\cite{Vojta-review,Vojta-Lee-prl}: the average 
activity density decays as power laws with continuously varying exponents as a 
function of the control parameter $p_\mathrm{d}$. A critical point, 
characterized by activated scaling logarithmic time decay 
($p_{\mathrm{d}_\mathrm{thr}}\approx 0.0402(1)$ in Fig.\ref{fig:decay}) 
separates the Griffiths phase from the active phase, marking the recovery of 
critical behavior at low dimension. 
Griffiths phases are a manifestation of rare-region effects:  
islands of localized activity are able to remain active for long times. Their 
relevance for both complex networks \cite{MAM-GP} and brain 
networks \cite{NatCommun} has been recently discussed in the literature.
Activity propagation models without signal integration 
yield such behavior in HMNs as they naturally lead to 
continuous phase transitions regardless of dimensionality 
constraints \cite{NatCommun}.
Remarkably, the $(2,1)$-model dynamics at low dimensions recovers here those 
fingerprints of criticality, in spite of being typically associated with 
discontinuous transitions at mean field.
Interestingly, upon approaching the threshold dimension from above, $D\to 2^+$, 
the discontinuous nature of the transition is rounded: 
although coexistence is genuinely recovered at very large times, activity is 
able to self-sustain even in the absorbing state for times potentially longer 
than any observation window.

\section{Discussion and Conclusions} 

A recent study showed that the $(2,1)$-process adopted here as a
paradigm for first-order phase transitions may show tricritical
behavior in certain families of ordered fractal lattices of dimension
$1<D<2$ \cite{Windus-Jensen09}. Such a finding implies that for each
family of ordered fractals, there exist a ``critical'' dimension
$1<D_\mathrm{c}<2$, below which the transition is continuous,
recovering the known behavior of one dimensional chains, and above
which the transition becomes discontinuous, anticipating the behavior
of pure two-dimensional lattices. In our study we have introduced
disorder in the topology and shown that finite-dimensional disordered
hierarchical modular networks of relevance in neuroscience {\it
  always} display continuous phase transitions for $D<2$. In fact even
square lattices ($D=2$) exhibit this behavior provided that disorder
is introduced, specifically in the form of quenched impurities
\cite{Villa14}.  Such results corroborate the conjecture that, due to
disorder, non-equilibrium systems with absorbing states do not sustain
first-order dynamical phase transitions for {\it any} $D\le 2$.  We
provided further evidence for this claim, proving its validity for
HMNs, and we focused on the relevance that such result may have for
neuroscience. Brain activity is known to exhibit critical-like
behavior, which would suggest its ability to sit constantly in the
vicinity of a continuous transition. We have shown that even if
realistic dynamic models lead to first-order phase transitions in the
mean-field approximation, in low-dimensional disordered systems such
transitions are rounded. A natural question arises whether the brain
actually is a low-dimensional network, provided that each single
neuron may have up to thousands of neighbors. The solution to this
apparent contradiction comes from the hierarchical organization of
brain connections. At the lowest scales, neurons are grouped in well
connected moduli which act as small worlds of diverging topological
dimension. At such scales, integrate-and-fire dynamics naturally
trigger coexistence and local discontinuous activations of moduli.  At
the largest scales, however, inter-moduli connectivity is very sparse
in order to maintain the volume of white-fiber matter bounded
\cite{Kaiser10}, allowing only for weak small-world effects \cite{Gallos}.  
Such connectivity patterns become finite dimensional, and discontinuous phase 
transitions are prohibited. 
Notice that the $D=2$ bound should not be read strictly
in real systems. We found that significant traits of quasi-critical
behavior appear even above $D=2$, suggesting a gradual rounding
phenomenon. Such systems will theoretically show discontinuous
transitions for large enough times, yet they are able to sustain
anomalous activity for typical time windows of experimental
observations. More detailed and realistic models of
neural dynamics could be provided. While the behavior presented here
is conserved for different choices of the parameters $(n,m)$,
realistic models supposedly include ingredients such as refractory
times, explicit time integration, inhibition and dependence on synapse
directedness. While such details are of primary importance in
correctly describing physiological aspects of brain activity, we
believe that our simple approach has the advantage of focusing on the
large-scale topology of the Human Connectome 
\cite{Sporns05,Hagmann}, in order to provide
insight about its large-scale behavior. An understanding of low-level
synaptic activity requires realistic neuron models and remains a
formidable task.

In conclusion, we have studied the properties of a family of dynamic
models of relevance in the description of neural activity in the
presence of signal integration. Although signal integration may be
responsible for the emergence of first-order phase transitions in
generic networks, we have shown that phase transitions are rounded in
finite dimensional hierarchical networks, eventually turning
continuous for $D\le 2$. Such finding is relevant in explaining the
observation of critical behavior in the brain at large scales, in
spite of the high degree of signal integration required to fire neuron
activity at small scale.

\section*{Acknowledgements}  
  
  We acknowledge support from the J. de Andaluc\'{i}a project of
  Excellence P09-FQM-4682 and from the Spanish MEC project
  FIS2009--08451.

\section*{References}
\providecommand{\newblock}{}

\end{document}